\documentclass[twocolumn,showpacs,preprintnumbers,amsmath,amssymb]{revtex4}

\usepackage{graphicx}
\usepackage{dcolumn}
\usepackage{bm}

\begin{document}

\title{Force-velocity relations for multiple-molecular-motor transport}

\author{Ziqing Wang$^{1,2}$}
\author{Ming Li$^{3}$}

\affiliation{${}^{1}$College of Science, Northwest A\&F
University, Yangling, Shaanxi 712100, China}

\affiliation{${}^{2}$Institute of Theoretical Physics, Chinese
Academy of Sciences, Beijing 100190, China}

\affiliation{${}^{3}$College of Physical Science, Graduate
University of Chinese Academy of Sciences, Beijing 100190, China}

\date{\today}

\begin{abstract}

A transition rate model of cargo transport by $N$
molecular motors is proposed. Under the assumption of steady state,
the force-velocity curve of multi-motor system can be derived from
the force-velocity curve of single motor. Our work shows, in the
case of low load, the velocity of multi-motor system can decrease or
increase with increasing motor number, which is dependent on the
single motor force-velocity curve. And most commonly, the velocity
decreases. This gives a possible explanation to some recent
experimental observations.

\end{abstract}


\maketitle

\section{Introduction}

The cargo transport by single cytoplasmic molecular motor has
been widely studied both experimentally \cite{block.03, car.05,
mallik.04, cop.97, cop.96} and theoretically \cite{fisher.07,
fisher.01, lie.07}. Cargoes \textit{in vivo}, however, are typically
transported by several motors \cite{gross.07} and sometimes even by
different kinds motors \cite{kural.05,muller.07}. By far little has
been known about the cooperativity of multiple motors during cargo
transport, and it is still an important and open research
subject. Especially, cargo transport by multiple processive
motors attracts much attention, since these motors can transport
cargoes over long distances without unbinding from the track, which
is convenient for experimental studies. Actually, experimental and
theoretical studies on such systems have been carried out in the
last decade. These works investigated systems of fluid-like cargoes
\cite{camp.06, bru.09, leduc.04}, or rigid or elastic cargoes
\cite{julicher.95, beeg.08, leduc.07, mouria.08, wang.05,
Igarashi.01}, in the existence of external load force. In the cases
of rigid or elastic cargo, the cargo can induce strong coupling
between motors, which is the focus of this article.

Recently, Klumpp \textit{et al.} proposed a transition rate model to
study the cooperative cargo transport by processive motors
\cite{Klumpp.05}. In their model, motors are supposed to share the
load force equally and have no other mutual interactions. By this
treatment, the author concluded that the velocity of the cargo
increases with the increasing motor number. Theoretical analysis
based on ratchet models also give the same results \cite{mouria.08,
wang.05, Igarashi.01}. Nevertheless, experiments have shown that the
cargo velocity is approximately independent of the number of motors
pulling the cargo \cite{beeg.08, leduc.07, Howard.89}. Very
recently, Shubeita \textit{et al.}'s experiment showed that increase
of kinesin number leads to slightly reduced cargo velocity
\cite{cell.08}. This result is out of expectation and contradicts
some theoretical results \cite{mouria.08, wang.05, Igarashi.01}, but
is supported by the simulation results \cite{gross.08}.

Most of the theoretical studies mentioned above fall into special
cases since they are dependent on the modeling of single motor
stepping. In this article we want to generally investigate the
dependence of the velocity of multi-motor system on the number of
motors pulling the cargo. For convenience, we suppose the motors can not detach from the track.
We proposed a steady state transition model of transporting cargo by
$N$ motors. Our calculation shows that the velocity of $N$
motors transport depends on the single motor's
force-velocity(F-V) relation, and especially in the case of low
load, the velocity of multi-motor system may decrease with
increasing motor number. This result provides a new explanation to
Ref. \cite{cell.08}, and our work predicts a general behavior of
multi-motor transport.

\begin{figure}[b]
\includegraphics[width=5.5cm]{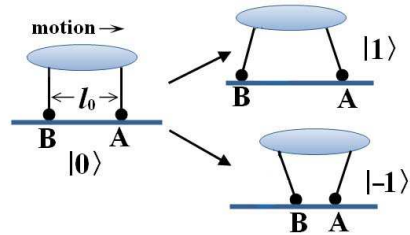}
\caption{\label{fig1} (Color online) A cargo is transported by two
motors A and B.}
\end{figure}

\section{Model and Results}
\subsection{Two-Motor case}

We first consider the
situation of two motors (A and B in Fig. 1) transporting a
cargo. The equilibrium distance between A and B is denoted as $l_0$.
If motor A takes a forward step, the distance between A and B
increases to produce a traction interaction between them \cite{jcb.93, epjb.98}, and the
cargo is assumed to step forward with a distance $d/2$, where $d$ is
the step size of the motor. This assumption is quite reasonable,
e.g., it has been shown experimentally that in the case of two
kinesins transporting a microtube, the step size of cargo is 4nm
which is half of a single motor's step size \cite{leduc.07}. When
motor B takes a forward step, resulting a repulsion interaction
between the two motors, and the cargo also take a forward step of
$d/2$. Here we focus mainly on the cases that the single motor
performs unidirected stepping, since it's a good enough description
of processive motors like kinesin.

States of the two-motor system can be specified by the ``effective
distances'' $S_{\mathrm{eff}}=S_{AB}-l_0$, $S_{AB}$ the real
distance between the two motors. So each state of the system can be
denoted by $| i \rangle$, where $i=S_{\mathrm{eff}}/d$ are integers.
When motor A takes a forward step, the system transits from state $|
i \rangle$ to state $| i+1 \rangle$. While motor B takes a forward
step, the system transits from state $| i \rangle$ to state $| i-1
\rangle$. Since the time for either motor performing the stepping
motions is much shorter than the dwell time(e.g., the 8-nm stepping
motion of kinesin occurs on the microsecond timescale while the
dwell time is always larger than millisecond \cite{car.05}. The
36-nm stepping motion of myosin-V occurs within a few millisecond,
far less than the dwell timescale of second \cite{cap.07}.), the two
motors can never step forward simultaneously. Therefore, the
transitions between states can be expressed as
\begin{equation}
|-n \rangle{\underrightarrow{~~\omega_{-n}^+~~} \atop
\overleftarrow{~\omega_{-n+1}^- }} \cdots
{\underrightarrow{\omega_{-2}^+} \atop \overleftarrow{\omega_{-1}^-
}}|-1 \rangle{\underrightarrow{\omega_{-1}^+} \atop
\overleftarrow{~\omega_0^- }} ~|0
\rangle~{\underrightarrow{~\omega_0^+} \atop
\overleftarrow{~\omega_1^- }} ~| 1 \rangle~
{\underrightarrow{~\omega_1^+} \atop \overleftarrow{~\omega_2^-
}}\cdots {\underrightarrow{\omega_{n-1}^+} \atop
\overleftarrow{~\omega_n^-~ }} |n \rangle
\end{equation}
The minus represents the distance between the two motors smaller
than $l_0$. The transition rates $\omega_i^{\pm}$ between states
depend on the external load force and the cargo-mediated force
between the two motors which will be discussed below. When stall
force is reached on either motor, the system gets into the extremity
states, $|n \rangle$ or $|-n \rangle$.

The cargo-mediated force exerted on either motor is quite intuitive,
i.e., when $i>0$, motor B exerts a resistance force $f$ on motor A
through the cargo, while motor A exerts an assistance force $-f$ on
motor B; and it is contrary when $i<0$. The magnitude of $f$ is
determined by the distance of the two motors and the stiffness of
motor linkage which connects the motor heads to the cargo as shown in Fig. 1. For different kinds of motor and cargo, the stiffness
of linkage are different. Here we take kinesin as an example, and
the methods can be extended to myosin-V directly, but it does not
apply to dynein because of their complexity and unclear stepping
behaviors \cite{mallik.04,cell.07}. The linkage of kinesin exhibited
an adequately linear behavior \cite{cop.97,cop.96}. In such cases,
the internal force between the two motors can be easily expressed as
$f=idk/2$, where $k$ is the linkage stiffness, and $i$ times $d$ is
the effective distance between the two motors and each motor shares
one half of the distance. When an external load $F$ is taken into
account, the total force borne by each motor should be $(F/2+f)$ or
$(F/2-f)$. Since the force-velocity relation $V_1(F)$ for single
motor transport has been widely studied both experimentally and
theoretically, one can easily know the step rates $R_1(F)=V_1(F)/d$
for either motor of the system. Then we can get the transition rates
$\omega_i^{\pm}$ in Eq. (1),
\begin{equation}
 \omega_i^{\pm}=R_1(F/2 \pm idk/2).
\end{equation}

Now we turn to the mean velocity of the two-motor system. Denoting
by $P_i$ the probability that the system is in state $|i \rangle$.
Here we are concerning the steady-state velocity of the system. The
steady-state solution of the process described by Eq. (1) can be
expressed as
\begin{equation}
 P_i=P_0 \prod_{j=0}^{i-1} \frac{\omega_j^+}{\omega_{j+1}^-}~~~
 \textrm{for}~~~(i>0),~~~\textrm{and}~~~P_{-i}=P_i.
\end{equation}
Considering the normalization $\sum \nolimits ^n_{i=-n}P_i=1$, $P_0$
satisfy
\begin{equation}
 P_0=\left[ 1 + 2\sum_{i=1}^{n}\prod_{j=0}^{i-1} \frac{\omega_j^+}{\omega_{j+1}^-} \right]^{-1}.
\end{equation}

In the case of linear spring linkage, when either of the two motors
takes a forward step, the cargo goes forward $d/2$. So the average
velocity of the cargo is then given by
\begin{eqnarray}
&&V_2(F)=\frac{d}{2}\sum_{i=-n}^n P_i(\omega_i^+ +
\omega_i^-)\nonumber\\
&&=\sum_{i=-n}^n P_i \frac{\left[ V_1(F/2 + idk/2) + V_1(F/2
-idk/2)\right]}{2}.
\end{eqnarray}

Defining $\widetilde{V}_2(F)\equiv V_2(2F)$. It's obvious that
$\widetilde{V}_2(F)<V_1(F)$ rigorously holds if the single motor F-V
curve is purely concave, which is followed by two facts: (1)
$V_2(F)<V_1(F)$ when the load $F$ is near zero, i.e, two-motor
transport is slower than single motor transport at low
load. (2) $V_2(F)>V_1(F)$ when $F$ is large, i.e., the two-motor
transport is generally faster. While single motor F-V curve is
purely convex, then $\widetilde{V}_2(F)> V_1(F)$ rigorously holds,
and two-motor transport is faster than single motor
transport in the whole range of $F$.

Most real single motor F-V curves, however, are usually a mixture of
concave and convex regions, so one can't intuitively know the
characteristics of the 2-motor F-V curve from the single motor F-V
curve, but can still get some insight of $V_2(F)$ when $F$ is near
zero. Roughly speaking, we have two typical categories of single
motor F-V curve.

\noindent \textit{Category A}: the single motor velocity is more
sensitive to resisting load than to assisting load (i.e., the F-V
region of assisting load is concave and much flatter than the region
of mediate resisting load, as illustrated by Fig. 2A. ),
$V_2(0)<V_1(0)$ may usually hold.

\noindent \textit{Category B}: the single motor velocity is more
sensitive to assisting load than to resisting load (as illustrated
by Fig. 2B), $V_2(0)>V_1(0)$ holds.

\noindent Therefore, for real single motor F-V curves, we can obtain
similar results as for purely concave and convex curves.

\begin{figure}
\includegraphics[width=8.5cm]{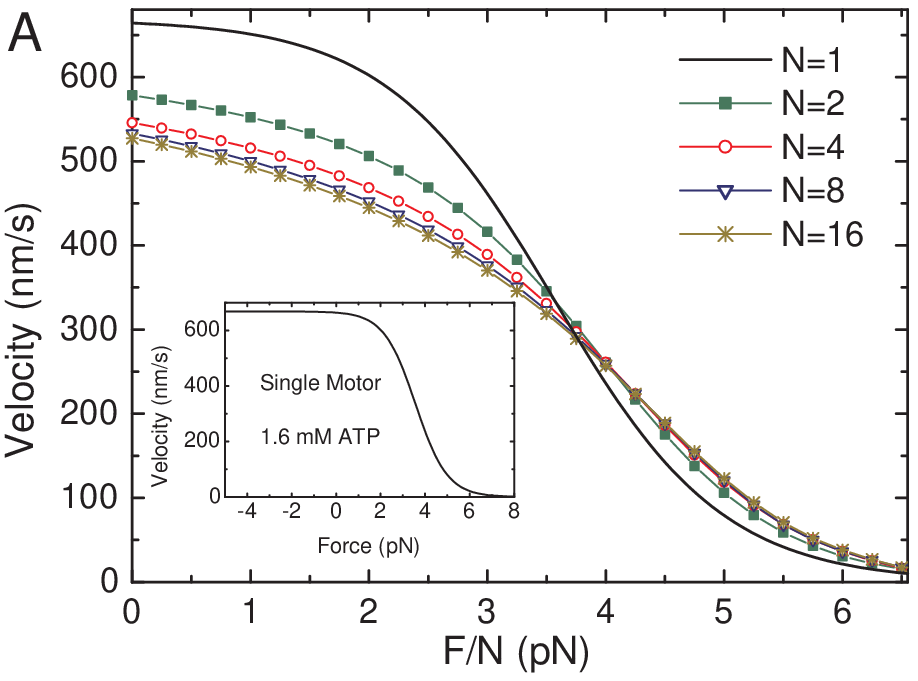}
\includegraphics[width=8.5cm]{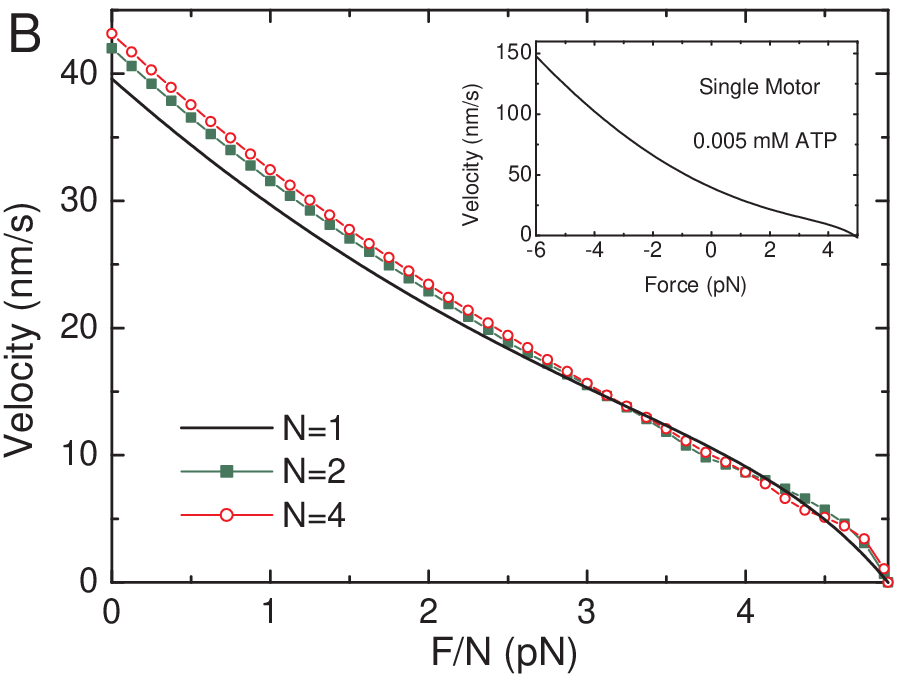}
\caption{\label{fig2} (Color online) The  $\widetilde{V}_N(F/N)$
curves of N-motor system derived from two typical single motor F-V
curves as shown in insets. The linkage stiffness is taken as
$k=0.3$pN/nm just for illustration. The value is adopted from the
experimental result of Ref. \cite{cop.96}. The curve in inset of (A)
is fitted from the experimental date of Ref. \cite{block.03} with
[ATP]=1.6 mM. The inset of (B) is adapted from theoretical results
of Ref. \cite{fisher.01} with [ATP]=5 $\mu$M.}
\end{figure}

For the nonlinear-spring motor linkage case, the step spacing of the
cargo varies. Eq. (5) seems not able to be used in this case. But
noticing that the average velocity of the cargo is equal to the
average velocity of either motor in the long time limit, so the
cargo velocity can be expressed as
\begin{equation}
V_2(F)=d \sum_i P_i \omega_i^+=d \sum_i P_i
\omega_i^-=\frac{d}{2}\sum_i P_i(\omega_i^+ + \omega_i^-),
\end{equation}
where the transition rates are $\omega_i^\pm= V_1(F/2 \pm f_i)/d$
and $f_i$ is the internal force between motors for state $|i
\rangle$. The last term of Eq. (6) is similar to Eq. (5).

Considering of backward steps, we set the forward step rate and
backward step rate for a single motor are $R_1^f(F)$ and $R_1^b(F)$
respectively with $V_1(F)=[R_1^f(F)-R_1^b(F)]d$, and their ratio is
$\varepsilon(F)=R_1^b(F) / R_1^f(F)$ which has been studied in Ref.
\cite{car.05}. If $V_1(F)$ and $\varepsilon(F)$ are given,
$R_1^f(F)$ and $R_1^b(F)$ can be known. The transition rates
$\omega_i^{\pm}$ in Eq.(1) are then
\begin{equation}
\omega_i^\pm=R_1^f(F/2 \pm f_i) + R_1^b(F/2 \mp f_i).
\end{equation}
One gets the probabilities $P_i$ of state $|i \rangle$ by Eq.(3) and
the average velocity of the two-motor system,
\begin{equation}
V_2(F)=\sum_{i=-n}^n P_i \frac{\left[ V_1(F/2 + f_i) + V_1(F/2 -f_i)
\right]}{2},
\end{equation}
which is the same form of Eq. (5).

\begin{figure}
\includegraphics[width=8cm]{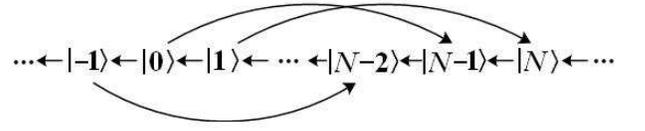}
\caption{\label{fig3} Transitions between states for N-motor system.
}
\end{figure}

\subsection{Multi-Motor case}
For a general N-motor system, we can
regard this system as the combination of a single motor and a
($N-1$)-motor subsystem. In order to conveniently describe the
model, we call this very single motor as motor A, and the
($N-1$)-motor subsystem as ``motor'' B. In the case of linear-spring
motor linkage, if one of the $N$ motors take a forward step, the
N-motor system will progress $d/N$. So the step size of motor A is
$d$, while the step size of ``motor'' B is $d/(N-1)$. Similar to the
two-motor system, we can express the states of the N-motor system by
the effective distance between motor A and ``motor'' B. Each state
denoted as $| i \rangle$, where $i$ is the value of the effective
distance between motor A and ``motor'' B divided by $d/(N-1)$, i.e.
$i=(N-1)S/d$, where $S$ is the effective distance between motor A
and ``motor'' B. When motor A takes a forward step, the N-motor
system will transit from state $| i \rangle$ to $| i+N-1 \rangle$,
while ``motor'' B takes a forward step, the system will convert from
state $| i \rangle$ to $| i-1 \rangle$. Transitions between the
states are shown in Fig. 3. The minus states represent that the
effective distance between A and B smaller than the equilibrium
distance. If the effective distance between A and B is $S$, then this distance shared by motor A is
$S(N-1)/N$ and shared by "motor" B is $S/N$, so the internal force between A and B is $f=idk/N$. Then the transition rates between the states can be given as
\begin{eqnarray}
\omega_i^+&=&V_1(F/N+idk/N)/d,   \nonumber\\
\omega_i^-&=&(N-1)V_{N-1}\left((N-1)F/N-idk/N\right)/d
\nonumber\\
&\equiv& (N-1)\widetilde{V}_{N-1}\left(F/N-idk/(N^2-N)\right)/d,
\end{eqnarray}
where the definition $\widetilde{V}_{N-1}(F/(N-1))\equiv V_{N-1}(F)$
is used. The steady state solution of the transition model shown in
Fig. 3 can be obtained if $V_1(F)$ is given and $V_{N-1}(F)$ is
known by the recursion of Eq. (9) and Eq. (10-11) given below. Any
one of the motors taking a forward step will make the system step
forward $d/N$. The average velocity of the N-motor system is then
given by
\begin{equation}
V_N(F)\equiv \widetilde{V}_N(F/N)=\frac{d}{N}\sum_i P_i(\omega_i^+ +
\omega_i^-).
\end{equation}
According to the network structure of Fig. 3, we can get $\sum_i
P_i\omega_i^-=(N-1)\sum_i P_i\omega_i^+$, the Eq. (10) can also be
expressed as
\begin{equation}
V_N(F)\equiv \widetilde{V}_N(F/N)=d\sum_i P_i ~\omega_i^+
=\frac{d}{N-1}\sum_i P_i ~\omega_i^-.
\end{equation}

If the number of motors $N$ is even, the above description can be
greatly simplified. Dividing these motors into two groups, each
group contain the same number of motors, $N/2$. Then this N-motor
system can be regarded as a two-big-motor system. If the F-V curve
of the (N/2)-motor system has already been known, we can easily
obtain the F-V curve for the N-motor system by the same method of
two-motor system. For convenience, we calculate the F-V curves for
this situation and show the results in Fig. 2.

Fig. 2A displays the F-V curves of multi-motor transport, as
well as the single motor F-V curve of \textit{Category A} which is
the most common case in many previous experimental and theoretical
studies. The calculation shows the velocity of multi-motor system
decreases with increasing motor number in the case of low load. This
offers a possible explanation to a recent experimental observation
\cite{cell.08}. We also notice that some experiments conclude that
the increase of motor number does't affect the transport
velocity of cargoes \cite{beeg.08,leduc.07,Howard.89}. This may be a
consequence of the fact that some F-V curves of single motor are not
far from linear, so all $V_N(0)(N=1,2\cdots)$ are almost equal,
i.e., multi-motor transport is not significantly slower than
single motor transport at low load.

One can also consider the consequence of motor detachment. Suppose $M$ motors adhering on a cargo.
The number $N$ of binding motors is no longer constant but varies with time between zero and $M$.
Therefore, the mean effective velocity of cargo transported by these $M$
motors can be expressed as weighted average of $V_N$ (e.g, Eq.(6) in \cite{Klumpp.05}),
\begin{equation}
V_{eff}^M(F)=\sum_{N=1}^M V_NP_N^M(F)
\end{equation}
where $P_N^M(F)$ is the force-dependent equilibrium binding probability of $N$ motors.
One can easily show that $V_{eff}^{M_1}(F)>V_{eff}^{M_2}(F)$ holds at low load
if $M_1<M_2$, by noticing that $V_N(F)$ decreases with $N$ at low load as shown in Fig. 2A.

\section{Discussions}
In this paper the F-V curve for cargo
transport by multiple motors has been discussed. We focused on
the linear-spring motor linkage case without considering the motor's
backward steps. The results show that the F-V curves of multi-motor
system depend on the F-V curve of single motor. Novel insights are
gained through our calculation, i.e., at low load, the velocity of
the multi-motor transport decreases with the increasing motor
number if the single motor F-V curve belongs to \textit{Category A},
and increases if the single motor F-V curve belongs to
\textit{Category B}. Our linear-spring motor linkage model can be
extended to the nonlinear-spring motor linkage case and also the
case of existence of backward steps. Results of both the latter
models are qualitatively consistent with the result of the former
model.

Our results contradict to earlier results which predict multi-motor
transport is faster than single motor transport
\cite{mouria.08,wang.05,Igarashi.01}. But the very recent experiment
supports our results, which shows that increasing of motor number
causes slight decrease of cargo velocity \cite{cell.08}. Ref.\cite{gross.08}
attributes the decreasing of
cargo velocity to the detachment of motors from filament. From our
results, even without motor detachment, the cargo velocity
can still decrease with increasing motor number. In fact, the
Fig. 2 of Ref. \cite{beeg.08} also shows a slight decrease of cargo
velocity with the increasing of motor numbers. There is another
difference between our results and the results of Ref.
\cite{gross.08}. In Ref. \cite{gross.08}, the simulation results
show that the multi-motor transport is slower than single motor
transport in low load case, but if the motor number is larger
than two, the cargo velocity will increase with the increasing of
motor number. Therefore, further experimental tests are needed, for
example, hopefully by the method of Ref. \cite{leduc.07}.

\section*{ACKNOWLEDGMENTS}
This project is supported by National Basic Research Program of
China (973 Program) 2007CB935903 and Research Foundation for Talents
of Northwest A\&F University.

\end{document}